\documentclass[%
 aip,
 jmp,%
 amsmath,amssymb,
 reprint,%
]{revtex4-1}

\usepackage{graphicx}
\usepackage{dcolumn}
\usepackage{bm}
\usepackage{color}

\begin{document}

\title{Coherence resonance in a network of FitzHugh-Nagumo systems: interplay of noise, time-delay and topology}

\author{Maria Masoliver}
\affiliation{Universitat Politecnica de Catalunya,
Colom 11, ES-08222 Terrassa, Barcelona, Spain} 

\author{Nishant Malik}
\affiliation{Department of Mathematics, Dartmouth College, Hanover, NH 03755, USA}

 \author{Eckehard Sch\"{o}ll}
 \affiliation{Institut f\"{u}r Theoretische Physik, Technische Universit\"{a}t Berlin, Hardenbergstr. 36, 10623 Berlin, Germany}

 \author{Anna Zakharova}
 \affiliation{Institut f\"{u}r Theoretische Physik, Technische Universit\"{a}t Berlin, Hardenbergstr. 36, 10623 Berlin, Germany}

\date{\today}

\begin{abstract}

We systematically investigate the phenomena of coherence resonance in time-delay coupled networks of FitzHugh-Nagumo elements in the excitable regime.  Using numerical simulations, we examine the interplay of noise, time-delayed coupling and network topology in the generation of coherence resonance.  In the deterministic case, we show that the delay-induced dynamics is independent of the number of nearest neighbors and the system size. In the presence of noise, we demonstrate the possibility of controlling coherence resonance by varying the time-delay and the number of nearest neighbors. For a locally coupled ring, we show that the time-delay weakens coherence resonance. For nonlocal coupling with appropriate time-delays, both enhancement and weakening of coherence resonance are possible.

\end{abstract}

\maketitle

\begin{quotation}

The FitzHugh-Nagumo system is a paradigmatic model which describes the excitability and spiking behavior of neurons. It has various applications ranging from biological processes to nonlinear electronic circuits. In the excitable regime under the influence of noise, this model exhibits the counterintuitive phenomenon of coherence resonance. It means that there exists an optimum intermediate value of the noise intensity for which noise-induced oscillations become most regular. We investigate coherence resonance in a network of delay-coupled FitzHugh-Nagumo elements with local, nonlocal and global coupling topologies. Networks with nonlocal topology are inspired by neuroscience, as they emulate the observation that strong interconnections between neurons are typical within a certain range while fewer connections exist at longer distances. We illustrate that the interaction between the network topology, the time-delay in the coupling, and the noise leads to a rich oscillatory dynamics. In particular, we demonstrate that the regularity of this dynamics is controllable, i.e., one can enhance or weaken coherence resonance by varying the coupling and delay time.

\end{quotation}

\section{Introduction}
\label{sec:1}

All natural processes are inevitably affected by internal and external random fluctuations, i.e., noise. Even a relatively low noise intensity can significantly influence the behavior of a dynamical system. In nonlinear systems noise can play a constructive role and give rise to new dynamic behavior, e.g., stochastic bifurcations, stochastic synchronization, or coherence resonance \cite{nsyn1,nsyn2,HU93a,nsyn3}. The counterintuitive effect of coherence resonance describes the non-monotonic behavior of the regularity of noise-induced oscillations in the excitable regime. This results in an optimum response in terms of the regularity of the oscillations for an intermediate noise strength.

In addition to noise, the presence of time-delay can essentially change the dynamics of a real-world system. Time-delay naturally arises in many processes, including population dynamics, chemical reactions, and lasers \cite{ERN09}. Interestingly, time-delay has not only been used to describe these processes but also to control them. For instance, when introduced in a nonlinear dynamical system it can control deterministic chaos \cite{SCH07}. Delay can control noise-induced oscillations as well, and consequently such effects as stochastic resonance and coherence resonance. A passive self-adaptive method for controlling noise-induced oscillations already exists,  delayed feedback previously used to control deterministic chaos forms the basis of this approach \cite{MAS02,JAN03, BAL04, SCH04b, MAS08}. Since then, several studies have been conducted on both excitable and non-excitable systems as well as on single and coupled oscillators \cite{HAU06, GEF14, PRA07, JUS09, FLU13}. These studies illustrate that delayed feedback effectively manipulates the properties of coherence resonance and adjusts the timescales of oscillations. Past studies have revealed that introduction of time-delayed feedback in a single system can control coherence resonance \cite{GEF15,JUS16}. In many systems, there are physical reasons for including time-delay in their modeling. For example, in neuroscience combining coupling with time-delayed feedback is a convenient approach to describe signal transmission in neuronal networks, i.e., the propagation delay of action potentials between neurons. Meanwhile, modeling studies have shown that presence of time-delay coupling can regulate the dynamics in networks, including stochastic synchronization in noise-affected systems and coupled lasers \cite{SCH08,HOE10b,FLU11a,SOR13, SAH17}.

The objective of this work is to investigate the interplay between noise, delay, and network topology of time-delay coupled neurons, where the FitzHugh-Nagumo model in the excitable regime represents the local dynamics of each neuron.  In particular, we are interested in the phenomena of  coherence resonance \cite{GEF15,HU93a,PIK97,HU00,JAN03,BAL04,SCH04b,LIN04,HAU06,AUS09, ROS09, ZAK10a,ZAK13,GEF14,BAL14,SEM15,SEM16}. Thus far, control of coherence resonance has been studied in single FitzHugh-Nagumo and in two coupled FitzHugh-Nagumo oscillators with time-delayed feedback \cite{HAU06}. In contrast, here we aim to investigate the control of coherence resonance in a network of delay-coupled FitzHugh-Nagumo oscillators.

The organization of the paper is as follows.  In Sec.~\ref{sec:2} we introduce the model and describe the behavior of a single FitzHugh-Nagumo oscillator. In Sec.~\ref{sec:3} we characterize the regimes of delay-induced oscillations in the deterministic case. Next in Sec.~\ref{sec:4}, we discuss the stochastic case but without the time-delay. We introduce different measures of coherence resonance and present an analysis of coherence resonance in a network of oscillators without the delayed coupling. We also explore the dependence of coherence resonance on the coupling parameters as well as on the bifurcation parameter.  Finally, in Sec.~\ref{sec:5} we investigate the interplay of noise, delayed coupling, and network topology.  We explore in detail how the time-delay and nearest neighbor coupling influence the coherence resonance. We conclude in Sec.~\ref{sec:6} with the summary of the results.

\section{Model}
\label{sec:2}

Throughout the paper, the model considered is a network of $N$ coupled FitzHugh-Nagumo oscillators. A FitzHugh-Nagumo oscillator is a minimalistic prototypical model of an excitable system \cite{FIT61a,NAG62}.  Excitable systems possess a single stable rest state and remain in the rest state unless perturbed by a sufficiently strong external input. Once perturbed, the system leaves the rest state and passes through the firing and the refractory states. The external driving has only a weak influence on the firing and refractory state\cite{LIN04}. Nonlinear dynamical systems exhibiting above properties have been proposed as models for neuronal spike generation. In neuroscience, the large excursion of the system's variables due to strong external perturbation (forcing the system to leave the rest state) is called a spike, and their occurrence as firing.  The excitability of a neuron can be classified into two categories namely, type I and type II.  Whereas type-I neurons undergo a saddle-node infinite period bifurcation, type-II neurons undergo a supercritical Hopf bifurcation \cite{LIN04, IZH00, RIN89, LEH15b}. A phenomenological description of this distinction also exists in the classical work of Hodgkin and Huxley \cite{HOD48}. The FitzHugh-Nagumo oscillator has been employed to model type-II neurons.  

The following set of equations describe a ring network of $N$ FitzHugh-Nagumo oscillators  \begin{gather}
\begin{split}
\epsilon \dot{u}_i &= u_i - \frac{u_i^3}{3} - v_i + \frac{\sigma}{2P} \sum_{j=i-P}^{i+P} [u_j(t - \tau) - u_i(t)] \\
\dot{v_i} &= u_i + a + \sqrt{2D}\xi_i(t), \hspace{0.5 cm} i=1,...,N 
\end{split}
\label{eq:FHN_1}
\end{gather} 
where $u_i$ and $v_i$  are dimensionless variables. The voltage-like variable $u_i$ allows for regenerative self-excitation through positive feedback, i.e., it is an activator variable; $v_i$ is a recovery-like variable and provides a slower negative feedback, i.e., it is an inhibitor variable.  The index $i$ stands for the node $i$ in the ring network of $N$ oscillators.  The time-scale parameter $\epsilon$  is usually much smaller than $1$ for neuronal models; here we set  $\epsilon = 0.01$. $P$ denotes the number of nearest neighbors to each side. For a ring, every node has the same number of connections; this gives rise to two limiting cases of local and global coupling, $P = 1$ and $P = (N-1)/2$ (for odd $N$), respectively. Note that for sufficiently large $N$, global coupling can be approximated by  $P = N/2$. When $1 < P < N/2$ we call it non-local coupling. Thus, $P$ acts as a control parameter for the topology of the underlying network.  $\sigma$ is the constant coupling strength and the coupling term has the form of classical diffusive coupling, i.e., the coupling vanishes if the variables $u_i$ and $u_j$ are identical.  $\tau$ is the propagation delay. $D$ stands for the noise intensity.  In this work, we use Gaussian white noise represented by $\xi(t)$ with  $\xi(t)\rangle = 0$  and $\langle \xi(t)\xi(t')\rangle = \delta(t-t')$ for $t \neq t'$ \cite{GAR02}. 
$a$ is the deterministic bifurcation parameter. A single FitzHugh-Nagumo system in the deterministic case ($D=0$) undergoes a supercritical Hopf bifurcation at $a = 1$. For $|a| < 1$ the system is in the oscillatory regime where the steady state is unstable and self-sustained oscillations are observed. For $|a| > 1$ the system is in the excitable regime and characterized by a locally stable steady state.

\section{Deterministic case: impact of time-delay}
\label{sec:3}

\begin{figure}
\centering
\includegraphics[width=\columnwidth]{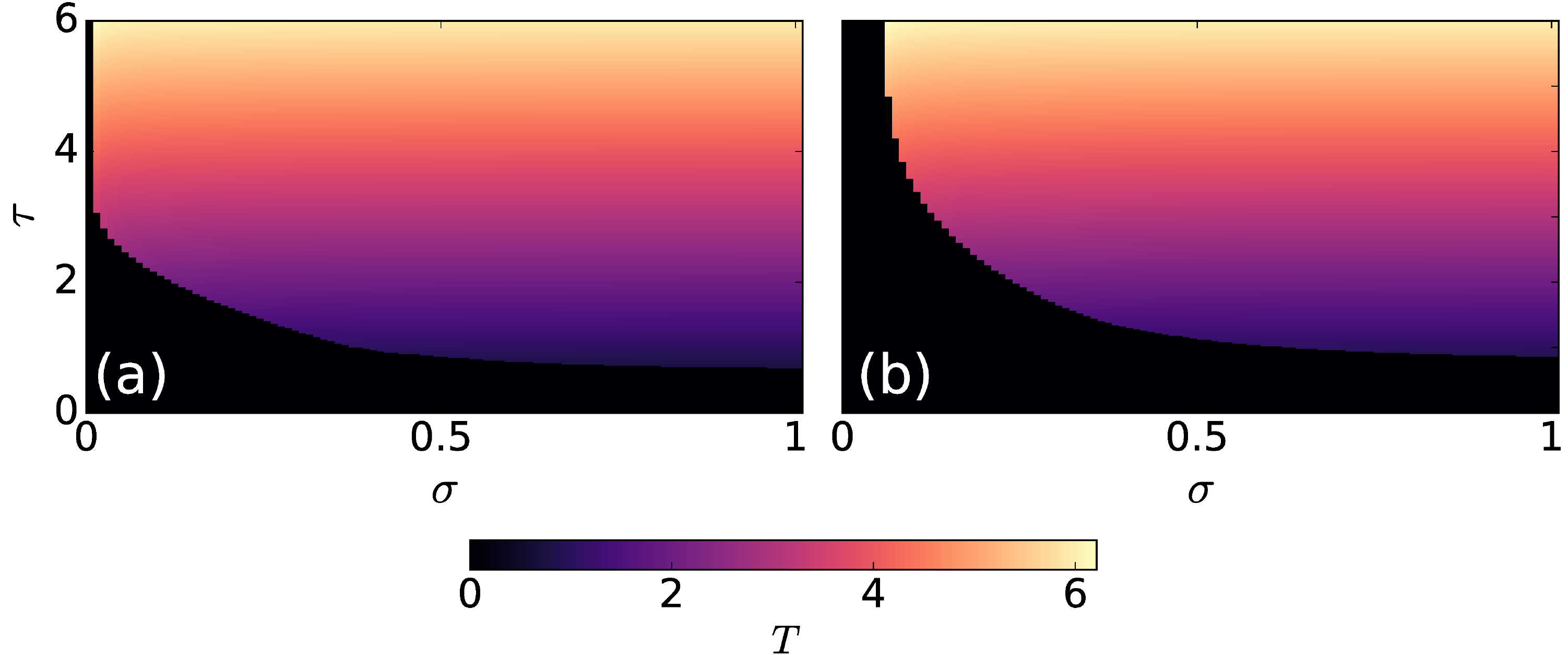}  
\caption{Regime of delay-induced oscillations in the ($\tau, \sigma$) plane for different values of the bifurcation parameter: (a) $a = 1.05$, (b) $a = 1.3$. The period of oscillations $T$ is color coded and corresponds to $T =  \tau + \delta$ with small $\delta>0$. The black region denotes absence of delay-induced oscillations. The initial history function corresponds to a spike for all oscillators. Other parameters: $\epsilon = 0.01$, $N$ = 100, $P = 4$, $D$=0.}
\label{fig:fig1}       
\end{figure}

To study the effect of delayed coupling on coherence resonance, the system must be in the parameter regime where no delay-induced oscillations exist. However, for certain time-delays and coupling strengths, delayed coupling induces self-sustained oscillations between two coupled FitzHugh-Nagumo oscillators, even when both oscillators are in the excitable regime (stable steady state) and there is no external noise applied \cite{SCH08, DAH08c}. A saddle-node bifurcation resulting in a pair of stable and unstable limit cycles generate these oscillations.\cite{SCH08} To identify parameter regimes where delay coupling induced oscillations are absent, we first study a ring network of $N$ FitzHugh-Nagumo oscillators in the deterministic regime by setting $D = 0$ in Eq.~(\ref{eq:FHN_1}). We numerically integrate Eq.~(\ref{eq:FHN_1}) for different values of $\tau$ and $\sigma$ and calculate the interspike interval or the oscillation period $T$ of synchronized oscillations. The results for $a=1.05$ and $a=1.3$ are shown in Fig.~\ref{fig:fig1}(a) and (b) respectively, where $T$ is plotted in the parameter space of coupling strength $\sigma$ and delay time $\tau$.  Fig.~\ref{fig:fig1}(a) with $a=1.05$ is closer to the Hopf bifurcation point, and we observe that further away from the bifurcation point (Fig.~\ref{fig:fig1}(b) with $a=1.3$) we require larger delay $\tau$ and coupling strength $\sigma$ to obtain delay-induced oscillations. The black region in Fig.~\ref{fig:fig1}(a) and (b) stands for the absence of delay-induced oscillations: this is the regime on which we focus in this work. 

For both Fig.~\ref{fig:fig1} (a) and (b) we use $P=4$, however, it can be shown that the region of delay-induced oscillations is independent of the ring topology and system size: Consider the delayed-coupling term in Eq.(\ref{eq:FHN_1}): 
$ \dfrac{\sigma}{2P}\sum\limits_{j=i-P}^{i+P}\left[u_j(t-\tau) -u_i(t)\right],\hspace{0.15 cm} j \neq i
$. Since delay-induced oscillations are synchronized, i.e., 
${u_1(t-\tau) = \cdots = u_N(t-\tau) \equiv u_{sync}(t-\tau)}$ and
${u_1(t) = \cdots = u_N(t) \equiv  u_{sync}(t)}$ where $u_{sync}$ is the synchronized solution,
we can simplify the delayed coupling term  as $\sigma \left[u_{sync}(t-\tau) -u_{sync}(t)\right]$.
 Rewriting Eq.(\ref{eq:FHN_1}) for the deterministic case, we find
\begin{gather}
\begin{split}
\epsilon \dot{u}_{sync} &= u_{sync} - \frac{u_{sync}^3}{3} - v_{sync} + \sigma \left[u_{sync}(t-\tau) -u_{sync}(t)\right] \\ 
\dot{v}_{sync} &= u_{sync} + a.
\label{sync}
\end{split}
\end{gather}
Observe that this equation is independent of both the nearest neighbor number $P$ and the number of oscillators $N$. Thus, the regime of delay-induced synchronized oscillations is independent of the ring topology and system size. 

%
%

\section{Coherence resonance}
\label{sec:4}

Pikovsky and Kurths \cite{PIK97} coined the term {\em coherence resonance} to characterize the emergence of relatively coherent oscillations in a FitzHugh-Nagumo system at an optimal noise intensity. Since then this phenomenon has been extensively studied in various nonlinear models. Several different measures exist in the literature for quantifying coherence resonance, such as the correlation time, the signal-to-noise-ratio, and the normalized standard deviation of the interspike interval.\cite{GEF14,HU93a,PIK97} In this work, we will use the last one. It is defined as   ${R=\displaystyle\frac{\sqrt{\langle{{t_{ISI}^2}}\rangle - {\langle{t_{ISI}}\rangle}^2}}	{\langle{t_{ISI}} \rangle}}$,
where $t_{ISI}$ is the time between two subsequent spikes and $\langle \cdots \rangle$ indicates the average over the time series.  A system undergoing coherence resonance will show a pronounced minimum in the value of $R$.\cite{PIK97} The above definition of $R$ is limited to characterizing coherence resonance for a single FitzHugh-Nagumo oscillator.  For a network of oscillators, coherence resonance can be measured by redefining $R$ as follows:  
\begin{equation}
R=\frac{\sqrt{\langle \overline{{t_{ISI}^2}}\rangle - {\langle \overline{t_{ISI}}\rangle}^2}}	{\langle\overline{t_{ISI}} \rangle}.
\label{eq:R_network}
\end{equation}
Where the over-line indicates the additional average over nodes.  We will refer to  ${\langle \overline{t_{ISI}}\rangle}$ as the period of the system $T$. Moreover, we refer to the period that the system shows under coherence resonance as the intrinsic period of the system and denote it by $T_{o}$.\\

\begin{figure}
\centering
\includegraphics[width=\columnwidth]{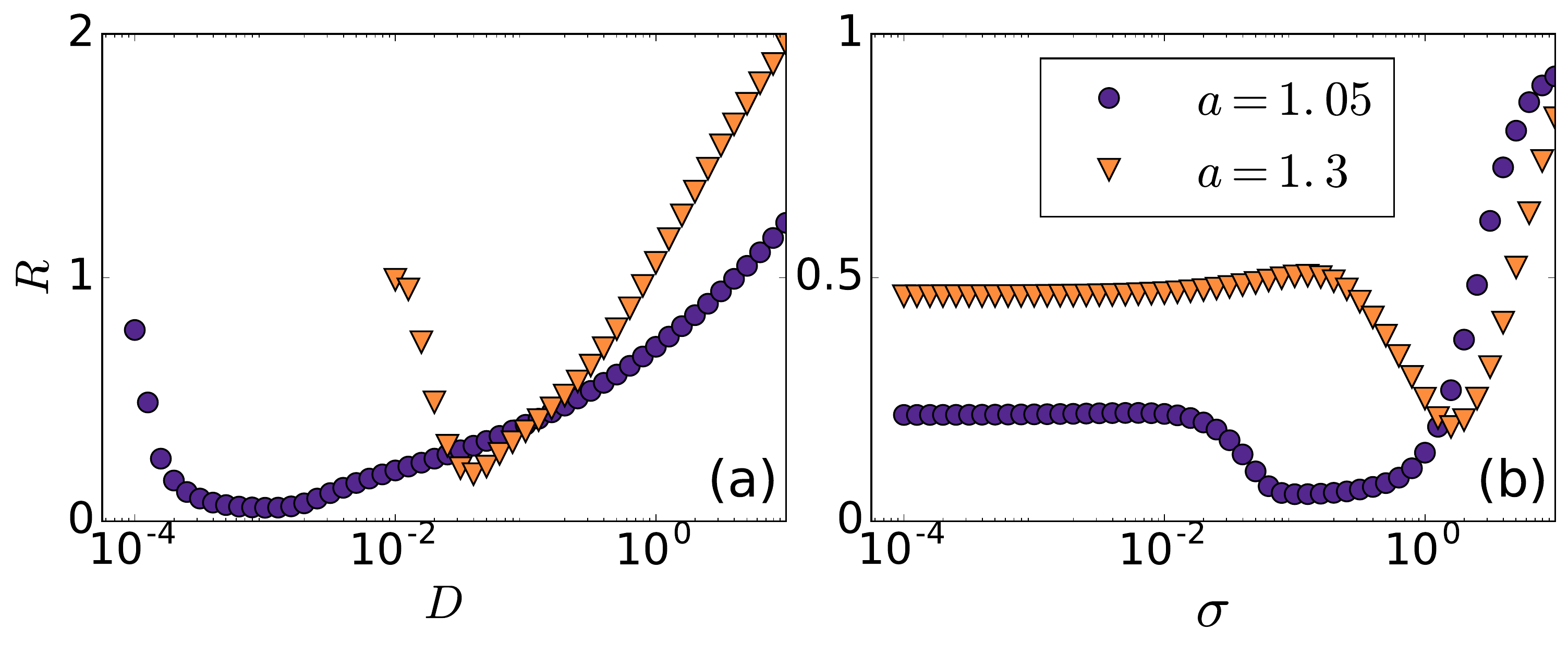} 
\caption{Normalized standard deviation of the interspike interval $R$ for two different values of the bifurcation parameter $a = 1.05$ (circles) and $a = 1.3$ (triangles): (a) for fixed coupling strength $\sigma = 0.1$ and varying noise intensity $D$; (b) for fixed noise intensity $D$=0.001 (for $a = 1.05$) and $D$=0.079 (for $a = 1.3$) and varying coupling strength $\sigma$. The results are obtained by integrating Eq. (\ref{eq:FHN_1}) over 10000 time units and then averaging over time, oscillators and realizations (for 20 simulations each). Note the logarithmic scale for the x-axis. Other parameters: $\epsilon = 0.01$, $P = 1$, $N = 100$, $\tau=0$.}
\label{fig:CR_P1}
\end{figure}

Next we study the role of noise intensity $D$ and coupling strength $\sigma$ in inducing coherence resonance in a network of locally coupled ($P$=1) FitzHugh-Nagumo oscillators without delay. We measure $R$ in two different parameter settings, first we increase $D$, keeping all parameters fixed and second we increase $\sigma$, keeping all the other parameters fixed.  The results are shown in Fig.~\ref{fig:CR_P1}(a) and (b); note that the x-axis is logarithmic. In  Fig.~\ref{fig:CR_P1}(a) both curves for $a=1.05$ and $a=1.3$ have a minimum, i.e.,  both cases show coherence resonance  at two different noise intensities $D$. It is worth noting here that if the system is closer to the Hopf bifurcation point, i.e., for $a=1.05$, it requires lower noise intensity for coherence resonance to occur. On the other hand, if the system is further away from the Hopf bifurcation point, i.e., for $a=1.3$, the system requires higher noise intensity. We observe $D = 0.001$ for the former, and  $D = 0.079$ for the latter case.  In Fig.~\ref{fig:CR_P1}(a) we have set $\sigma = 0.1$, $P = 1$ and $N = 100$. To study the effects of coupling strength on the above observed coherence resonance, we measure $R$ as $\sigma$ is varied in two different parameter settings: first, for $a=1.05$ and $D = 0.001$, and second, for $a=1.3$ and $D = 0.079$. The results are plotted in Fig.~\ref{fig:CR_P1}(b). We observe for the case $a=1.05$ and $D = 0.001$ that coherence resonance is enhanced when  $0.1 \leq \sigma <1$. In Fig.~\ref{fig:CR_P1}(b) we have set $P = 1$ and $N = 100$. Several other works have also shown that coherence resonance can be enhanced by choosing appropriate coupling strengths. For example in Refs.~\onlinecite{HU00} and \onlinecite{KWO02} it was shown that some choices of coupling strength increase coherence resonance in an array of non-identical FitzHugh-Nagumo oscillators; in Ref.~\onlinecite{BAL14} a similar feature was observed in the case of weighted scale-free networks.


\begin{figure}
\includegraphics[width=\columnwidth]{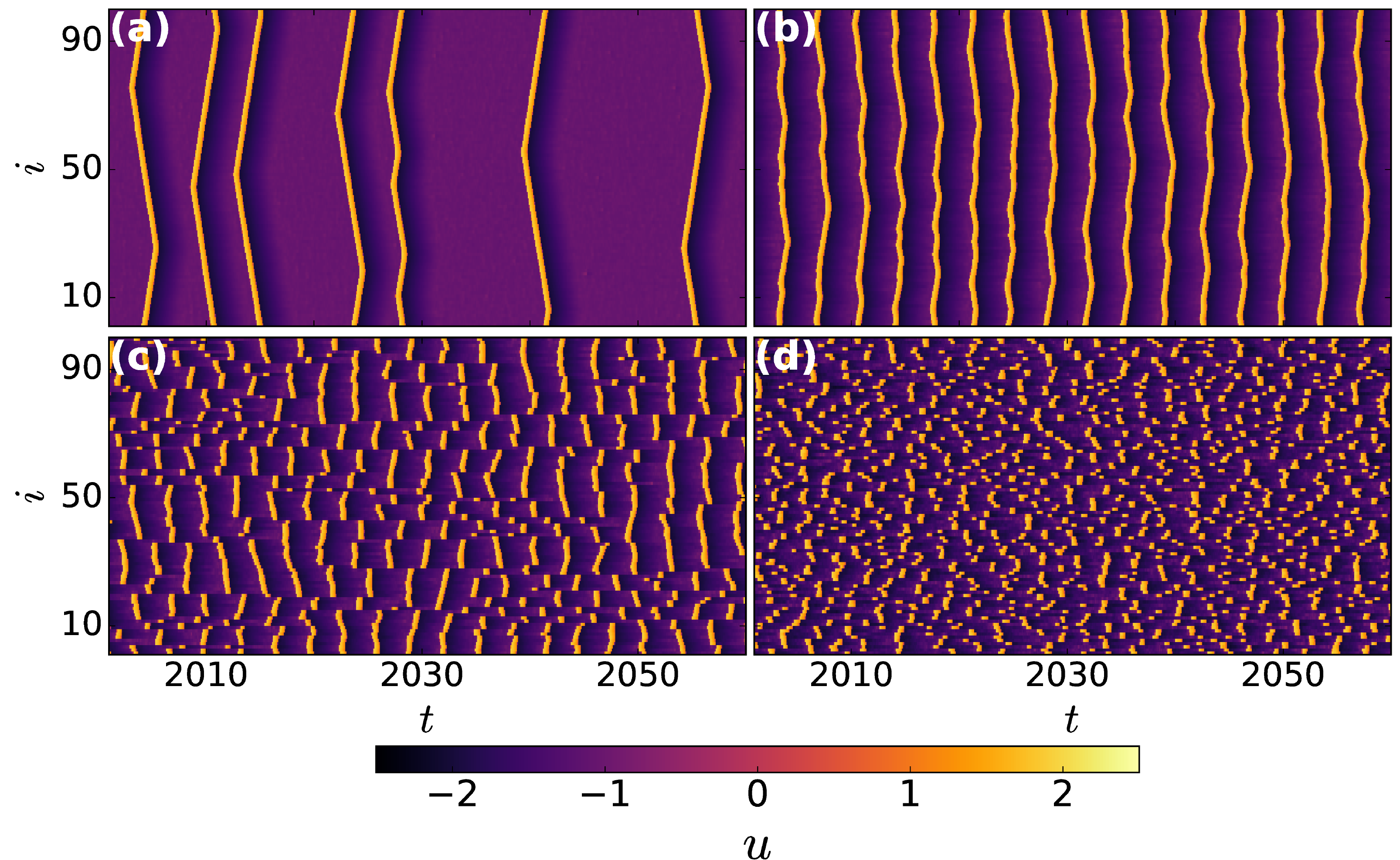} 
\caption{Space-time plots for $a = 1.05$ at different noise intensities (a) $D = 0.00012$, (b) $D = 0.001$, (c) $D = 0.005$, (d) $D = 0.05$. Other parameters: $\epsilon=0.01$, $\tau=0$, $P=1$, $N=100$, $\sigma = 0.1$.}
\label{fig:stp_P1}      
\end{figure}

To visualize the above observations, in Fig.~\ref{fig:stp_P1}(a-d), we depict space-time plots for different noise intensities. For  small noise $D = 0.00012$, in Fig.~\ref{fig:stp_P1}(a) we observe that the system is spiking irregularly but still it is in a highly synchronized state. This implies that the deterministic coupling dominates the system dynamics for small noise intensities. As the noise intensity is increased to an optimal value $D = 0.001$, in Fig.~\ref{fig:stp_P1}(b) we observe highly regular synchronous spiking, this is the parameter regime where we observe coherence resonance. Once noise exceeds its optimal value, in Fig.~\ref{fig:stp_P1}(c) the system exhibits cluster synchronization i.e., several clusters of synchronously spiking oscillators are formed. When external noise is further increased, in Fig. \ref{fig:stp_P1}(d) cluster synchronization disappears and each node oscillates individually, driven by its own noise.

\begin{table}[t!]
\centering
\begin{center}
\begin{tabular}{ ||c c c c c c||} 
\hline
 \multicolumn{6}{||c||}{\textit{Non-local and global coupling, $\tau=0$}} \\
 \hline\hline
						&$P = 1$	& $P = 4$ &$P = 12$ & $P = 25$  & $P = 50$\\ 

$D_{o}$ 				& 0.001			 	& 0.001 & 0.0008 & 0.0008 	& 0.0008\\ 
$T_{o}$	 		&3.53 & 3.51 & 3.53 & 3.61 	& 3.62\\ 
$R_{o} $ 			&0.06				 		& 0.04 & 0.032 & 0.029	& 0.029\\
\hline
\end{tabular}
\end{center}
\caption{Values of parameters $D$, $T$ and $R$ at coherence resonance when $P$ (number of nearest neighbors) is varied. 
Other parameters: $\epsilon=0.01$, $a=1.05$, $\sigma = 0.1$ and $N = 100$. }
\label{table:3}
\end{table}

So far we have only discussed coherence resonance in a ring network with $P=1$ (every node has exactly two neighbors). To explore the impact of $P$, we fix  $\sigma = 0.1$ and $a = 1.05$, and study the system with four different values of $P$, namely  $P = 4$, $P = 12$, $P = 25$, and $P = 50$ (all to all connected network). For each case we calculate the noise intensity $D_o$ for which we observe coherence resonance. Also we evaluate the corresponding values of  $R$ and $T$, and denote them as $R_o$ and $T_o$ respectively. We summarize our findings in Table.~\ref{table:3}. We notice that as $P$ is increased $D_o$ and $R_o$  decrease, indicating that we require lower noise intensity to observe stronger coherence resonance for higher $P$.  


\section{Interplay of topology and delayed coupling}
\label{sec:5}

In this section, we explore the effects of topology and delayed coupling on noise-induced oscillations. In a single FitzHugh-Nagumo oscillator, a time-delay can either enhance or suppress coherence resonance.\cite{BAL04} If the delay is $\tau = nT_{o}$ then for an integer $n$  coherence resonance increases, whereas for half-integer $n$ it is weakened.\cite{BAL04}  In contrast to Ref.~\onlinecite{JAN03,BAL04} where a single FitzHugh-Nagumo oscillator is analyzed, we study a network of $N$ delay-coupled FitzHugh-Nagumo oscillators. Therefore, we will not only illustrate the influence of time-delay but also the topology.

We will investigate the effects of $\tau = \frac{1}{2}T_{o}$ and $\frac{1}{3}T_{o}$ on a network of oscillators described by $P = 1$, $P = 4$, $P = 25$ and $P = 50$. We divide our results into two parts based on the types of coupling; the first part is about the locally-coupled ring and the second about the non-locally and globally coupled ring. 
It is important to note that $T_{o}$ refers to the period of the network with particular topology when it exhibits coherence resonance. Hence, for each $P$ we have a different $T_{o}$. 

\subsection{Locally coupled ring}

\begin{figure}
\includegraphics[width=\columnwidth]{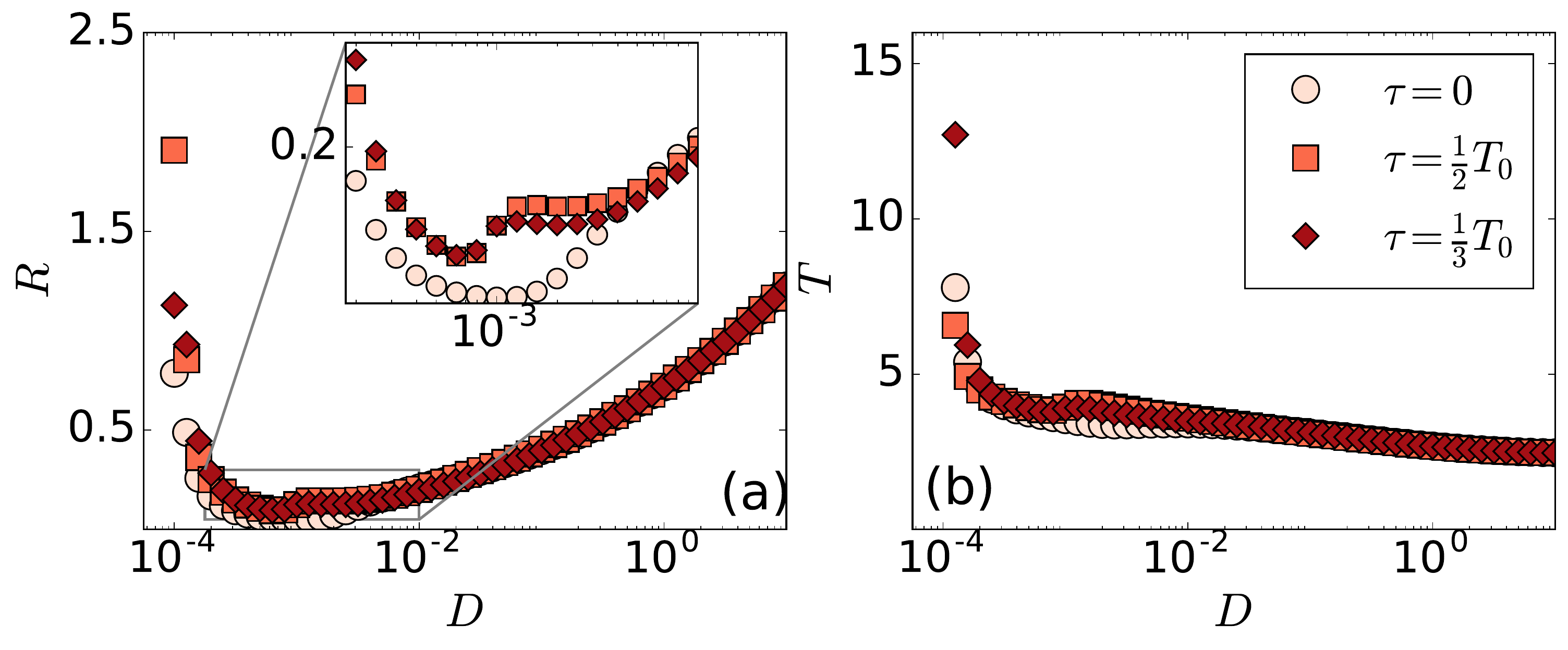} 
\caption{(a) $R$ vs. noise intensity $D$, (b) period of oscillations $T$  vs.  noise intensity $D$ 
for different time-delays in the case of a locally coupled ring, i.e., $P=1$. The inset in panel (a) shows a zoom of the indicated region. 
Other parameters: $\epsilon=0.01$, $a = 1.05$, $N = 100$ and $\sigma = 0.1$.
}
\label{fig:control_CR_1} 
\end{figure}

First we study $P=1$, i.e., a locally coupled ring. We plot $R$ and the period of oscillations $T$ vs. noise intensity $D$ in Fig. \ref{fig:control_CR_1} (a) and (b), respectively. The corresponding values of $R_o$,  $D_o$ and $T_o$ are listed in the first column of Table \ref{table:5} and \ref{table:6}. Comparing the values of $R_o$ for $P=1$ without time-delay (Table I) and in the presence of time-delay (Table II and III), one can see that for $\tau = \frac{1}{2}T_{o}$ and $\frac{1}{3}T_{o}$ coherence resonance is slightly weakened. Also, coherence resonance occurs at the same noise intensity $D_o$, with almost the same minimal value $R_o$  for both delays. Once the noise is sufficiently large, it overtakes the dynamics of the network, and delayed-coupling does not play a role any longer. As shown in Fig.~\ref{fig:control_CR_1}(a), for both $\tau$ values the system undergoes a local minimum in $R$, further increase in noise intensity leads to a monotonic increase in $R$.

\begin{table}[h!]
\centering
\begin{center}
\begin{tabular}{ ||c c c c c||} 
\hline
 \multicolumn{5}{||c||}{\textit{Delay-coupled network $\tau = \frac{1}{2} T_{o}$}} \\
 \hline\hline
										& $P = 1$ &$P = 4$ & $P = 25$  & $P = 50$\\ 

$D_{o}$ 							 	& 0.0006 & 0.0004 & 0.00025 	& 0.0002\\ 
$T_{o}$ 	& 3.85 & 3.66 & 3.75 	& 3.8\\ 
$R_{o} $ 							 		& 0.094 & 0.036 & 0.01	& 0.007\\
\hline
\end{tabular}
\end{center}
\caption{Values of parameters $D$, $T$ and $R$ at coherence resonance when $P$ (number of nearest neighbors) is varied. 
Other parameters: $\epsilon=0.01$, $a=1.05$, $\sigma = 0.1$ and $N = 100$.}
\label{table:5}
\end{table}
\begin{table}[h!]
\centering
\begin{center}
\begin{tabular}{ ||c c c c c||} 
\hline
 \multicolumn{5}{||c||}{\textit{Delay-coupled network $\tau = \frac{1}{3} T_{o}$}} \\
 \hline\hline
										& $P = 1$ &$P = 4$ & $P = 25$  & $P = 50$\\ 

$D_{o}$ 							 	& 0.0006 & 0.0004 & 0.0005 	& 0.0006\\ 
$T_{o}$	& 3.79 & 3.96 & 4.18 	& 4.26\\ 
$R_{o} $ 							 		& 0.096 & 0.092 & 0.127	& 0.159\\
\hline
\end{tabular}
\end{center}
\caption{Values of parameters $D$, $T$ and $R$ at coherence resonance when $P$ (number of nearest neighbors) is varied. 
Other parameters: $\epsilon=0.01$, $a=1.05$, $\sigma = 0.1$ and $N = 100$. }
\label{table:6}
\end{table}


\subsection{Non-locally and globally coupled ring}

\begin{figure}
\centering
\includegraphics[width=\columnwidth]{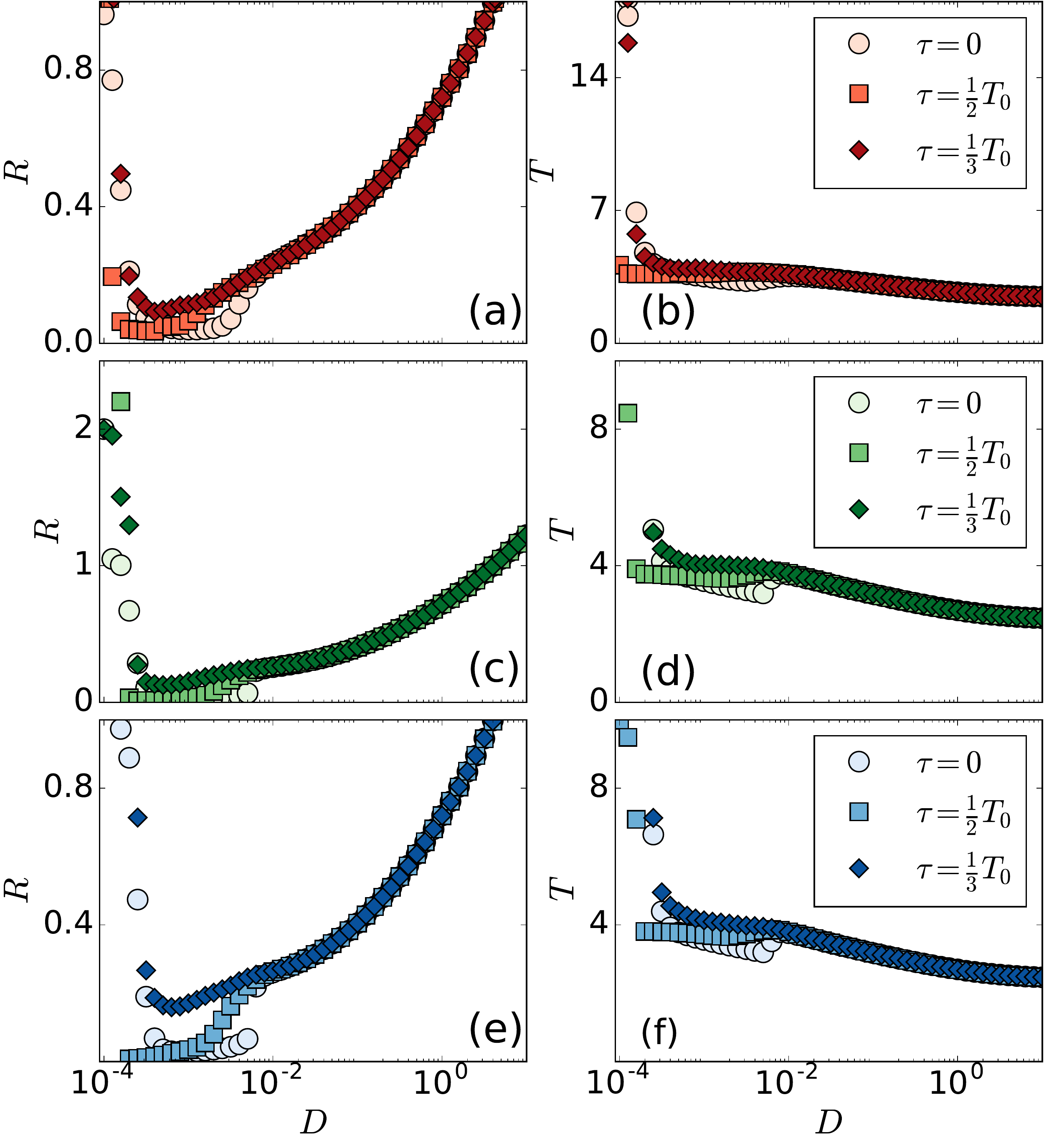} 
\caption{Same as Fig.~\ref{fig:control_CR_1} for a non-locally coupled ring: (a),(b) $P = 4$; (c),(d) $P = 25$; (e),(f) globally coupled ring ($P = 50$). Panels (a), (c) and (e) show $R$ vs. noise amplitude $D$, panels (b), (d) and (f) show oscillation period $T$ vs. $D$  for different time-delays (see legend for specific values). Other parameters: $\epsilon=0.01$, $a = 1.05$, $N = 100$ and $\sigma = 0.1$.}
\label{fig:control_CR}      
\end{figure}

Now we study a non-locally coupled ($1<P<50$) and a globally coupled ($P=50$) ring network. The non-locally and globally coupled ring lead to two different outcomes: while coherence resonance is enhanced for  $\tau = \frac{1}{2}T_{o}$, it is weakened for $\tau = \frac{1}{3}T_{o}$, compared with the undelayed case (see the values of $R_o$ from tables I and II,III). The corresponding numerical results are displayed in column two to four in Table~\ref{table:5} and Table~\ref{table:6} and plotted in Fig. \ref{fig:control_CR} (a-f).

For  $\tau = \frac{1}{2}T_{o}$ coherence resonance is strengthened as $P$ increases, the strongest coherence resonance being observed for global coupling (see Table~\ref{table:5} or Fig.~\ref{fig:control_CR} (a,c)). In contrast, for  $\tau = \frac{1}{3}T_{o}$, coherence resonance is weakened with increasing $P$  (see Table~\ref{table:6} or Fig.~\ref{fig:control_CR} (e)). Also, when  $\tau = \frac{1}{2}T_{o}$, in Table~\ref{table:5} we observe that coherence resonance occurs at smaller noise intensities compared to $P=1$. It should be noted that  for both $\tau = \frac{1}{2}T_{o}$ and  $\tau = \frac{1}{3}T_{o}$, the values of $T_{o}$ are higher than one observed for $\tau =0$ in  Table~\ref{table:3}.

In Table~\ref{table:6} we note that increasing $P$ leads to increasing $R_0$, i.e., the irregularity of the motion increases.  This particular observation suggests that for $\tau = \frac{1}{3}T_o$ delayed-coupling induces irregularity in the oscillations.  It destabilizes the time interval between successive spikes and nodes. Hence, it increases the range of variation of the period $T$ under the change of noise strength  (also see Fig. \ref{fig:control_CR} (b), (d) and (f)).

\section{Summary and conclusions}
\label{sec:6}

We have systematically studied coherence resonance in a network of delay-coupled FitzHugh-Nagumo oscillators and have presented a detailed analysis of rich dynamics emerging due to the interactions between noise, time-delayed coupling, and topology. First, we  demonstrated that in a ring network of deterministic FitzHugh Nagumo delay-coupled oscillators, the regions of delay-induced oscillations are independent of the number of nearest neighbors $P$  and system size. Furthermore,  we showed that these regions depend on the bifurcation parameter $a$ and they grow when $a$ is further away from the Hopf bifurcation. 

Next, we considered the stochastic case without time-delay, i.e., $D \neq 0$ and $\tau = 0$. We observed that coherence resonance can be enhanced or weakened by the coupling strength $\sigma$.  With increasing $\sigma$ we have found a minimum in the normalized variance of the interspike interval  $R$, i.e., coherence resonance is strengthened. We studied coherence resonance for two different values of the bifurcation parameter, viz. $a = 1.05$ and $a = 1.3$ (with $P = 1$), and found that larger noise intensity is required to observe coherence resonance in the latter case. Moreover, for  $a = 1.3$ higher coupling strength is needed for coherence resonance. That is why changing the system from $a = 1.05$ to $a = 1.3$ increases the range of no-delay-induced oscillations; coherence resonance requires higher noise and coupling. Therefore, keeping the system at $a = 1.05$ is better suited to study the system, as it is more sensitive to noise, the number of nearest neighbor $P$, and coupling strength. 

On including non-zero time-delay into the system, several new features emerged. We explored the system with two different time-delays, namely $\tau = \frac{1}{2}T_{o}$ and $\tau = \frac{1}{3}T_{o}$. 
Whereas for a locally coupled ring ($P=1$), delay-coupling weakens coherence resonance for both values of $\tau$, in the case of a non-locally  ($1<P<50$)  and globally ($P=50$) coupled ring we found different results depending on $\tau$. For $\tau = \frac{1}{2}T_{o}$ an enhancement of coherence resonance is observed, while for $\tau = \frac{1}{3}T_{o}$ coherence resonance is weakened. This is due to the influence of an \textit{indirect} coupling: node $i$ is directly coupled to $2P$ nodes, additionally it is  \textit{indirectly} coupled to $(2P)^2$ neighbors  with a delayed coupling of $2\tau$. Hence, for $\tau = \frac{1}{2}T_{o}$ the total propagation delay is equivalent to $T_{o}$ and for $\tau = \frac{1}{3}T_{o}$ to $\frac{2}{3}T_{o}$, leading to the enhancement or weakening of coherent dynamics.

\end{document}